\documentclass[preprint,aps,showpacs]{revtex4}

\usepackage{graphicx}

\begin{document}
\title{Non-exponential decay via tunneling in tight-binding lattices and
the optical Zeno effect} \normalsize
\author{Stefano Longhi}
\address{Dipartimento di Fisica and Istituto di Fotonica e Nanotecnologie del CNR,
Politecnico di Milano, Piazza L. da Vinci 32, I-20133 Milano,
Italy}


%
\bigskip
\begin{abstract}
\noindent An exactly-solvable model for the decay of a metastable
state coupled to a semi-infinite tight-binding lattice, showing
large deviations from exponential decay in the strong coupling
regime, is presented. An optical realization of the lattice model,
based on discrete diffraction in a semi-infinite array of
tunneling-coupled optical waveguides, is proposed to test
 non-exponential decay and for the observation of an optical analog
of the quantum Zeno effect.
\end{abstract}


\maketitle

\newpage
The understanding and control of the decay process of an unstable
quantum state has long been a subject of debate in different areas
of physics. Though an exponential law is known to be a good
phenomenological fit to many decay phenomena, quantum mechanics
ensures that the survival probability $P(t)$ is definitely {\it
not} exponential at short and long times (see, e.g.,
\cite{Winter61,Ghirardi78,Nakazato96}). In particular, at short
times $P(t)$ always shows a parabolic decay, i.e. $dP/dt
\rightarrow 0$ as $ t \rightarrow 0$. These universal features
have been extensively investigated in some specific models
describing the tunneling escape of a particle through a potential
barrier \cite{Winter61,Razavy,Dijk99}, or in the framework of the
exactly-solvable Friedrichs-Lee Hamiltonian
\cite{Prigogine91,Facchi99,Kofman00,Facchi01,Kofman01}, which
describes the decay of a discrete state coupled to a continuum.
The short-time features of the decay process have attracted much
attention because they can lead, under certain conditions, to
either the deceleration (Zeno effect) or the acceleration
(anti-Zeno effect) of the decay by frequent observations of the
system (see, e.g., \cite{Kofman00,Facchi01,Chiu77} and references
therein). Evidences of non-exponential decay features at short
times and the observation of the related Zeno and anti-Zeno
effects have been reported in recent experiments on quantum
tunneling of trapped sodium atoms in accelerating optical lattices
\cite{Willkinson97}. Similar effects have been proposed to occur
for quantum tunneling
in analogous macroscopic systems, such as Josephson junctions \cite{Barone04}.\\
In this Letter a novel and exactly-solvable model of
non-exponential decay of an unstable state tunneling-coupled to a
tight-binding lattice is presented. A simple and experimentally
accessible realization of the model, based on discrete diffraction
of photons in an array of optical waveguides
\cite{Christodoulides03}, is
proposed along with an optical analog of the quantum Zeno effect.\\
To set our model in a general context, we consider a semi-infinite
lattice described by the tight-binding Hamiltonian [Fig.1(a)]:
\begin{equation}
H_{TB}= - \hbar \sum_{n=1}^{\infty} \Delta_n \left( | n \rangle
\langle n+1| + | n+1 \rangle \langle n| \right),
\end{equation}
\begin{figure}
\includegraphics[scale=0.8]{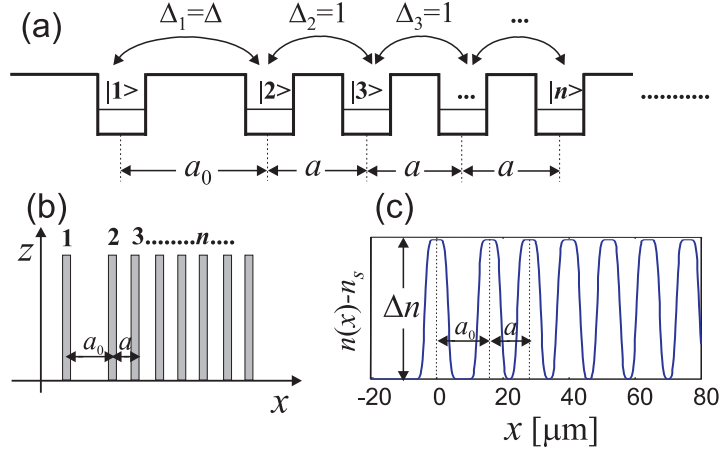} \caption{
(a) The semi-infinite tight-binding lattice model. (b) Optical
realization of the tight-binding model based on an array of
coupled optical waveguides. (c) Refractive index profile
$n(x)-n_s$ of the waveguide array used in the numerical
simulations (parameter values are: $n_s=2.138$, $\lambda=1.55 \;
\mu$m, $\Delta n=2.4 \times 10^{-3}$, and $a=12 \; \mu$m).}
\end{figure} 
where $|n\rangle$ ($n \ge 1$) is the state localized at the $n$-th
site of the lattice and $\Delta_n$ is the hopping amplitude
between adjacent sites $|n\rangle$ and $|n+1\rangle$. We assume
that for $n \geq 2$ the lattice is periodic so that, after a
rescaling of time $t$, we may assume $\Delta_n=1$ for $n \geq 2$.
The boundary site $|1\rangle$ is then coupled to the periodic
lattice by a hopping amplitude $\Delta_1=\Delta$, which is assumed
to be smaller than $1$. The tight-binding Hamiltonian (1) has been
often used as a simple model to describe coherent transport
properties
 and tunneling phenomena in different
physical systems, including semiconductor superlattices
\cite{Holthaus}, arrays of coupled quantum dots
\cite{Nikolopoulos04}, Bose-Einstein condensates in optical
lattices \cite{Trombettoni01}, and arrays of optical waveguides
\cite{Christodoulides03,Pertsch99}. In particular, model (1) can
be derived from the continuous Schr\"{o}dinger equation
\begin{equation}
i \hbar \frac{\partial \psi}{\partial t}=-\frac{\hbar^2}{2 m}
\frac{\partial^2 \psi}{\partial x^2}+V(x) \psi,
\end{equation}
with a potential $V(x)=\sum_{n=0}^{\infty} V_{w}(x-x_n)$
describing a semi-infinite chain of identical symmetric quantum
wells $V_{w}(x)$ [$V_w(-x)=V_w(x)$ and $V_w(x) \rightarrow 0$ for
$x \rightarrow \infty$], placed at distances $x_{n+1}-x_n=a$ for
$n \geq 2$ and $x_2-x_1=a_0>a$ [see Fig.1(a)]. If the individual
potential well $V_w(x)$ supports a single bounded mode $\varphi
(x)$ of energy $E$ and if tunneling-induced coupling of adjacent
wells is weak, Eq.(2) can be reduced to the discrete model (1) by
means of a tight-binding \cite{Trombettoni01,Ablowitz03} or a
variational \cite{Peall} analysis. After expanding the state
$|\psi\rangle$ of the system as $|\psi\rangle=\sum_n c_n(t)
\exp(-iE t / \hbar) |n \rangle$, where $|n\rangle=\varphi (x-x_n)$
is the localized state at the $n$-th well in the chain, in the
nearest-neighbor approximation from Eq.(2) one can derive the
following equations of motion for $c_n$:
 \begin{eqnarray}
i \dot c_1 & = & -\Delta c_2 \;, \;\;  i \dot c_2= -c_3-\Delta c_1 \; , \nonumber \\
 i \dot c_n & = & -(c_{n+1}+c_{n-1}) \;\; {\rm for} \; n \geq 3,
\end{eqnarray}
where $\Delta \simeq [\int dx \; \varphi (x-a_0)V_w(x) \varphi
(x)]/[\int dx \; \varphi (x-a)V_w(x) \varphi (x)]$ is the
normalized hopping amplitude between states $|1\rangle$ and
$|2\rangle$. For $\Delta=0$, i.e. for $a_0/a \rightarrow \infty$,
the site $|1\rangle$ is decoupled from the other lattice sites and
if the system is initially prepared in state $|1\rangle$ it does
not decay; as $\Delta$ is increased, tunneling escape is allowed
and state $|1\rangle$ becomes metastable. The limits $\Delta
\rightarrow 0$ and $\Delta \rightarrow 1$ correspond to the weak
and strong coupling regimes, respectively. The occupation
probability of site $|1\rangle$ at time $t$ is given by
$P(t)=|c_1(t)|^2$. Following Gamow's approach to quantum tunneling
decay \cite{Razavy}, the 'natural' decay rate $\gamma_0$ of state
$|1\rangle$, which would correspond to an exponential decay law
$P(t)=\exp(-\gamma_0 t)$, can be readily calculated  by looking
for complex energy eigenfunctions of $H_{TB}$ with outgoing
boundary conditions (Gamow's states), yielding:
\begin{equation}
\gamma_{0} = 2 \Delta^2 (1-\Delta^2)^{-1/2}.
\end{equation}
However, the exponential decay law turns out to be incorrect,
especially in the strong coupling regime $\Delta \rightarrow 1$
where it fails to reproduce the exact decay law at {\it any} time
scale. According to Ref.\cite{Facchi01}, one can introduce an {\it
effective} decay rate $\gamma_{eff}(t)$ by the relation
$\gamma_{eff}(t)=-(1/t) {\rm ln |c_1(t)|^2} $, so that any
deviation of $\gamma_{eff}(t)$ from $\gamma_0$ is a signature of
non-exponential decay. In addition, the eventual intersection
$\gamma_{eff}(t)=\gamma_0$ rules the transition from Zeno to
anti-Zeno effects for repetitive measurements \cite{Facchi01}. In
order to determine the {\it exact} law for the survival
probability $P(t)$, one has to calculate the eigenfunctions of (1)
and construct a suitable superposition of them corresponding, at
$t=0$, to a particle localized in the well $|1\rangle$, i.e. to
$c_n(0)=\delta_{n,1}$. The tight-binding Hamiltonian (1) has a
continuous spectrum of eigenfunctions \cite{note1} which can be
calculated by separation of variables and correspond to
$c_n(t)=u_n(Q) \exp[i \Omega(Q)t]$, where $\Omega(Q)=2 \cos Q$ is
the dispersion curve of the tight-binding lattice band, $-\pi <Q<
\pi$ varies in the first Brillouin zone, and:
\begin{eqnarray}
u_1 & = & \Delta(1+r)/(2 \cos Q),   \\
u_n & = & \exp[-iQ(n-2)]+r \exp[iQ(n-2)] \; (n \geq 2). \nonumber
\end{eqnarray}
In Eq.(5), $r=r(Q)$ is the reflection coefficient for Bloch waves
at the boundary of the semi-infinite lattice and reads explicitly:
\begin{equation}
r(Q)=-\frac{\Delta^2-2 \cos Q \exp(iQ)}{\Delta^2-2 \cos Q
\exp(-iQ)}.
\end{equation}
To study the decay process, we construct a superposition of the
eigenstates, $c_n(t)=\int_{-\pi}^{\pi} dQ \; F(Q) u_n(Q) \exp[i
\Omega(Q)t]$, where the spectrum $F(Q)$ is determined by the
initial conditions $c_n(0)=\delta_{n,1}$. Using an iterative
procedure that will be described in detail elsewhere, one can show
that the searched spectrum is given by $F(Q)=-(2 \pi \Delta)^{-1}[
\Delta^2 \exp(iQ)-2 \cos Q]/[\Delta^2-1-\exp(2iQ)]$. Therefore the
{\it exact} decay law for the occupation amplitude of site
$|1\rangle$ is given by:
\begin{equation}
c_1(t)  =  \frac{1}{2 \pi} \int_{-\pi}^{\pi} dQ \; \exp(2 i  t
\cos Q) \frac{1-\exp(-2iQ)}{1+\alpha^2 \exp(-2iQ)},
\end{equation}
where $\alpha \equiv (1-\Delta^2)^{1/2}$. The short-time decay of
$|c_1(t)|^2$ is obviously parabolic; the long-time behavior of
$c_1(t)$ can be calculated by use of the method of the stationary
phase, yielding the oscillatory power-law decay
\begin{equation}
c_1(t) \sim \frac{1}{\sqrt \pi} \frac{1-\alpha^2}{(1+\alpha^2)^2}
\frac{1}{ t^{3/2}} \cos(2 t - 3 \pi /4) \; \; {\rm as} \; t
\rightarrow \infty.
\end{equation}
In order to extract the  exponential decay part from $c_1(t)$,
after setting $z=\exp(iQ)$ it is worth rewriting Eq.(7) as an
integral in the complex plane:
\begin{equation}
c_1(t)=\frac{1}{2  \pi i} \oint_{\sigma} dz \; \exp \left[ i t
\left(z+\frac{1}{z}\right)  \right] \frac{z^2-1}{z(z^2+\alpha^2)},
\end{equation}
where the contour $\sigma$ is the unit circle $|z|=1$. The
integral (9) can be evaluated by use of the residue theorem. Note
that, for $\Delta=1$ there is only one singularity at $z=0$, and
from residue theorem one obtains:
\begin{equation}
c_1(t)=(1/t)J_1(2 t),
\end{equation}
which shows that, in the strong coupling regime, the decay greatly
deviates from an exponential law at {\it any} time scale. For
$\Delta<1$, there are three singularities, at $z=0$ and $z= \pm i
\alpha$, internal to the contour $\sigma$. The residue associated
to the singularity $z=-i \alpha$ yields an exponentially-decaying
term, whereas the sum of residues at $z=0$ and $z=i \alpha$ yields
a bounded function $s(t)$, which can be written as a Neumann
series. Precisely, one can write:
\begin{equation}
c_1(t)=\sqrt{\mathcal{Z}} \exp (-\gamma_0 t /2)+s(t),
\end{equation}
where $\gamma_0$ is the natural decay rate as given by Gamow's
theory [Eq.(4)], $\sqrt{\mathcal{Z}} \equiv (\alpha^2+1)/(2
\alpha^2)$, and
\begin{equation}
s(t)=J_0(2
 t)+\left( 1+\frac{1}{ \alpha^2} \right) \left[\frac{1}{2} \sum_{l=-\infty}^{\infty}  \frac{J_l(2
t)}{\alpha^{l}} -\sum_{l=0}^{\infty} \frac{J_{2l}(2
t)}{\alpha^{2l}} \right]
\end{equation}
\begin{figure}
\includegraphics[scale=0.6]{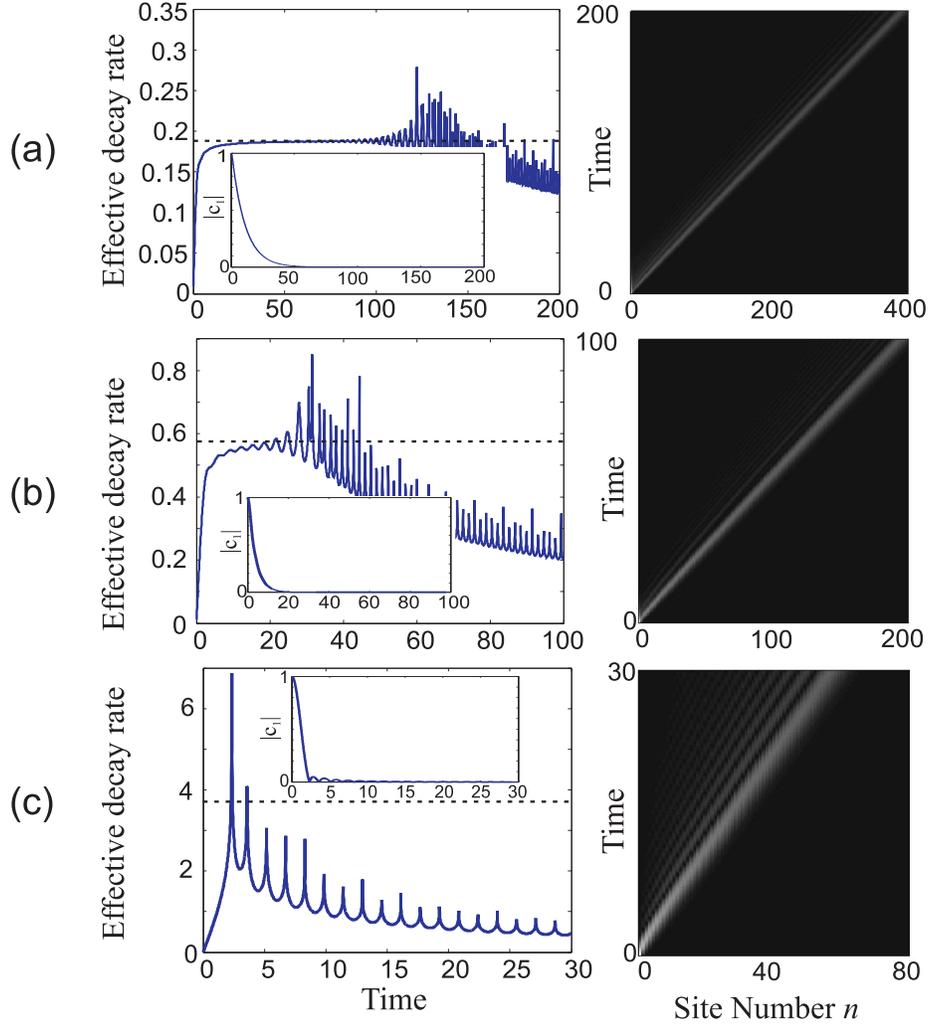} \caption{
Left: Behavior of the effective decay rate $\gamma_{eff}$ and
amplitude $|c_1(t)|$ (insets) versus time. Right: Grey-scale image
of $|c_n(t)|$. In (a), $\Delta=0.3$; in (b), $\Delta=0.5$; in (c),
$\Delta=0.9$. The horizontal dashed curves are the natural decay
rate $\gamma_0$.}
\end{figure} 
 is the correction to the exponential decay term. The
decomposition (11) is meaningful in the weak coupling regime
($\Delta \rightarrow 0$) since, in this limit, one can show that
the  contribution $s(t)$ is small and of order $\sim \Delta^2$.
The appearance of non-exponential features in the decay dynamics
when approaching the strong coupling limit is clearly shown in
Fig.2, where the numerically-computed behavior of the effective
decay rate $\gamma_{eff}(t)$ is shown for a few values of
$\Delta$, together with the temporal evolution of amplitudes
$|c_n(t)|$. The appearance of strong oscillations in the
$\gamma_{eff}(t)$ curve when the coupling strength increases is a
clear signature of an oscillatory decay dynamics which sets in
{\it even} at intermediate time scales. Consider now the case of
projective measurements of state $|1\rangle$ at time intervals
$t=\tau$. In the weak coupling limit, where the decay deviates
from an exponential law solely at short and long times,
deceleration of the decay (Zeno effect) occurs for $\tau<\tau^*$,
where $\tau^*$ is the smallest root of the equation
$\gamma_{eff}(\tau^*)=\gamma_0$ \cite{Facchi01}; for instance, for
parameter values of Fig.2(a) one has $\tau^* \sim 85$ . In the
strong coupling regime [Fig.2(c)], the decay is highly
oscillatory, and acceleration of the decay (anti-Zeno effect) may
be observed for a value of $\tau$ close to e.g. the first peak of
$\gamma_{eff}$, where $\gamma_{eff}(\tau)$ is larger than
$\gamma_0$; for instance, for parameter values of Fig.2(c)
anti-Zeno effect may be observed for $\tau \sim 2.34$. In this
case, repetitive observations correspond to suppression of the
oscillatory tails
in the decay process.\\
\begin{figure}
\includegraphics[scale=1]{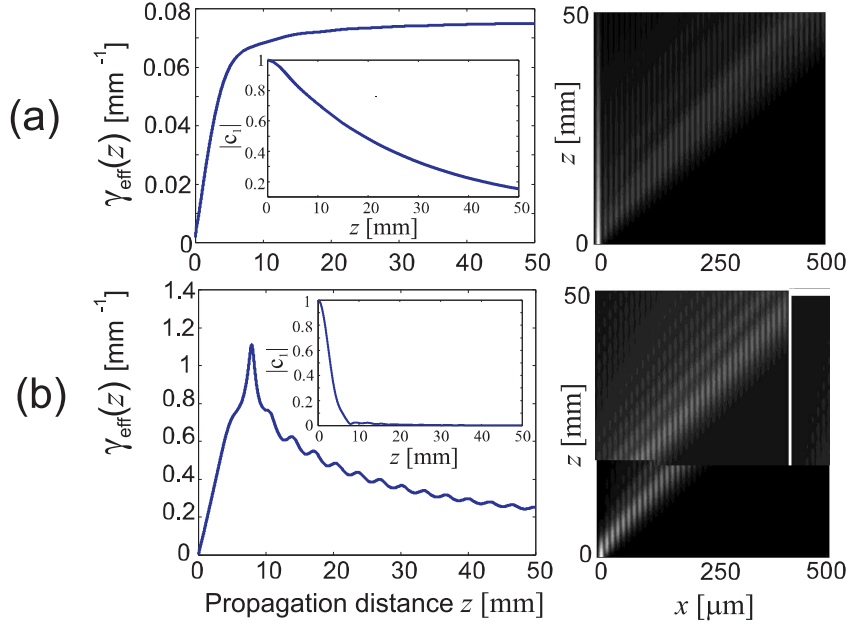} \caption{
Tunneling decay dynamics in a $L=50$-mm-long semi-infinite
waveguide array (left column) and corresponding discrete
diffraction patterns (right column). (a) Weak coupling regime
[$a=12 \; \mu$m and $a_0=16 \; \mu$m, corresponding to $\Delta
\sim 0.28 $]; (b) strong coupling regime [$a=12 \; \mu$m and
$a_0=12.5 \; \mu$m, corresponding to $\Delta \sim 0.86$].}
\end{figure} 
\noindent
 Physical realizations of the tight-binding model
(1) are provided by electron transport in a chain of
tunneling-coupled semiconductor quantum wells \cite{Holthaus} or
by discrete diffraction of photons in a semi-infinite array of
tunneling-coupled optical waveguides, where the temporal variable
$t$ of the quantum problem is mapped into the spatial propagation
coordinate $z$ along the array  [Fig.1(b)]. Here we consider in
detail the latter optical system since it shows several
advantages: (i) Visualization of the tunneling dynamics is
experimentally accessible \cite{Pertsch99,Trompeter06}, and a
quantitative measure of light decay can be done by e.g. NSOM
techniques \cite{Campillo03}; (ii) Preparation of the system on
state $|1\rangle$ is simply realized by initial excitation of the
boundary waveguide by a focused laser beam; (iii) Light
diffraction experiments in waveguide arrays have successfully
confirmed the reliability of the tight-binding model
\cite{Christodoulides03,Pertsch99}; (iv) Transport of photons
instead of charged particles (e.g. electrons) avoids the
occurrence of dephasing or many-body effects, making
waveguide-based optical structures an ideal laboratory for the
observation of several analogs of coherent quantum dynamical
effects (see, e.g. \cite{Trompeter06}).  Beautiful optical analogs
of Bloch oscillations
\cite{Christodoulides03,Pertsch99,Trompeter06}, Landau-Zener
tunneling \cite{Trompeter06}, adiabatic stabilization of atoms in
strong fields \cite{Longhi05}, and coherent control of quantum
tunneling \cite{Vorobeichik03},
have been indeed reported in recent optical experiments.\\
Light propagation in the waveguide array is described by Eq.(2) in
which the temporal variable $t$ is replaced by the spatial
propagation coordinate $z$, $\hbar=\lambda / (2 \pi)$ is the
reduced wavelength of photons, $m=n_s$ is the refractive index of
the array substrate, $V(x) \simeq n_s-n(x)$, and $n(x)$ is the
array refractive index profile (see, e.g.,
\cite{Vorobeichik03,Longhi05}). As an example, Fig.3 shows the
discrete diffraction patterns and corresponding behavior of light
trapped in waveguide $|1\rangle$ as obtained by a numerical
analysis of Eq.(2) using a standard beam propagation method with
absorbing boundary conditions \cite{Vassallo96}; initial condition
corresponds to excitation of waveguide $|1\rangle$ in its
fundamental mode, i.e. $\psi(x,0)=\varphi(x)$.
\begin{figure}
\includegraphics[scale=0.9]{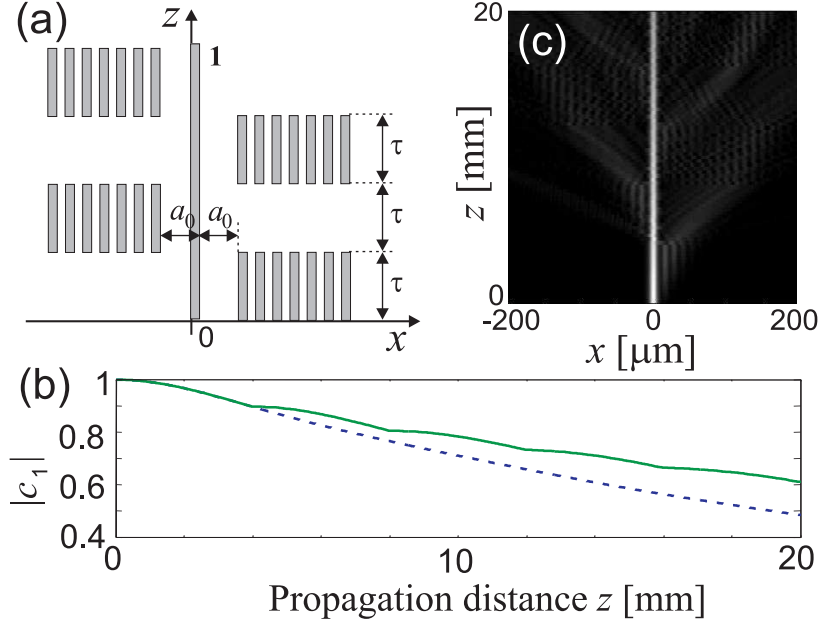} \caption{
(a) Schematic of a waveguide array for the observation of the
optical Zeno effect. (b) Numerically-computed behavior of mode
amplitude $|c_1|$ trapped in waveguide $|1\rangle$ (solid curve)
versus propagation distance in a $L=20$-mm-long array for $\tau=4$
mm, $a_0=16 \; \mu$m, and $a=12 \; \mu$m. The dashed curve is the
behavior corresponding to Fig.3(a). (c) Grey-scale discrete
diffraction pattern along the array.}
\end{figure} 
\noindent The refractive index profile of the semi-infinite array
used in the simulations is plotted in Fig.1(c) for parameter
values which typically apply to lithium-niobate waveguides
\cite{Longhi05}. Note that, as $\Delta$ is increased,
non-exponential features are clearly visible. However, as compared
to the tight-binding results, the peaked structure of
$\gamma_{eff}(t)$ obtained from the continuous model (2) is
smoothed [compare e.g. Fig.2(c) and Fig.3(b)]. In order to
reproduce the optical analog of the quantum Zeno effect in the
waveguide system, one can adopt the array configuration shown in
Fig.4(a), in which a straight waveguide $|1\rangle$ is
periodically coupled, at equally-spaced distances $z=\tau$, to
semi-infinite arrays of finite length $\tau$ placed on alternating
sides of the waveguide. At each section where the lateral arrays
end, light trapped in the interrupted waveguides is scattered out
and solely a negligible fraction of it will be re-coupled into the
waveguides at the next section of the array. Therefore, at planes
$z=\tau, 2 \tau, 3 \tau, ...$ one can assume, at first
approximation, that a collapse of the state $\psi(x,z)$ into the
fundamental mode $\varphi(x)$ of waveguide $|1\rangle$ occurs,
thus simulating the 'wavepacket collapse' of an ideal quantum
measurement. An example of deceleration of the decay via tunneling
 in the alternating array, analogous to the quantum Zeno effect,
 is shown in Figs.4(b) and (c).\\
In conclusion, an exactly-solvable model for the tunneling escape
dynamics of a metastable state coupled to a tight-binding lattice
has been presented, and its optical realization -including an
optical analog of the quantum Zeno effect- has been proposed in an
array of tunneling-coupled optical waveguides.


\begin{thebibliography}{}
\bibitem{note1}
For $\Delta \leq 1$, localized states at the lattice boundary do
not exist.
\end{thebibliography}
\end{document}